\documentclass[english,prl,twocolumn,superscriptaddress,showpacs,letterpaper]{revtex4}

\usepackage{ae} 
\usepackage[T1]{fontenc}
\usepackage[ansinew]{inputenc}
\usepackage{amsmath}
\usepackage{amssymb}
\usepackage{graphicx}
\usepackage{color}
\usepackage{epsfig}
\usepackage{booktabs}

\newcommand{\mx}{x}

\newcommand{\ph}{{\rm ph}}
\newcommand{\mir}{{\rm mir}}
\newcommand{\cut}{{\circlearrowright\hspace{-3mm}*\hspace{0.2mm}}}
\newcommand{\figm}{{\includegraphics[scale=0.5]{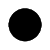}}}
\newcommand{\figf}{{\includegraphics[scale=0.5]{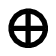}}}
\newcommand{\figF}{{\includegraphics[scale=0.5]{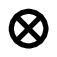}}}
\newcommand{\figb}{{\includegraphics[scale=0.5]{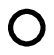}}}
\newcommand{\figp}{{\includegraphics[scale=0.5]{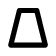}}}

\newcommand{\beq}{\begin{equation}}
\newcommand{\eeq}{\end{equation}}
\newcommand\beqa{\begin{eqnarray}}
\newcommand\eeqa{\end{eqnarray}}
\newcommand\bea{\begin{array}}
\newcommand\eea{\end{array}}

\newcommand{\im}{{\rm Im}\;}

\def\XXint#1#2#3{{\setbox0=\hbox{$#1{#2#3}{\int}$}
\vcenter{\hbox{$#2#3$}}\kern-.5\wd0}}

\newcommand{\nn}{\nonumber}

\newcommand{\COMMENT}[1]{}

\newcommand{\neqa}{\nonumber\end{eqnarray}}
\newcommand{\la}[1]{\label{#1}}

\renewcommand{\d}{\partial}

\newcommand{\<}{{\langle}}
\renewcommand{\>}{{\rangle}}

\newcommand{\re}{\relax{\rm I\kern-.18em R}}

\newcommand{\rb}{\right)}
\newcommand{\lb}{\left(}

\def\su2{{SU(2)}}

\def\<{\langle}
\def\>{\rangle}

\def\i2{\frac{i}{2}}

\usepackage{varioref}
\usepackage{makeidx}
\makeindex

\usepackage[english]{babel}
\begin{document}


\title{Exact AdS/CFT spectrum:\\  Konishi dimension at any coupling}

\author{ Nikolay Gromov}
\affiliation{DESY Theory, Hamburg, Germany \& II. Institut f\"{u}r Theoretische Physik Universit\"{a}t, Hamburg, Germany \&\\ St.Petersburg INP, St.Petersburg, Russia }
\author{Vladimir Kazakov}
\affiliation{Ecole Normale Superieure, LPT,  75231 Paris CEDEX-5, France $\&$ l'Universit\'e Paris-VI, Paris, France  $\&$ Institut Universitaire de France  }
\author{Pedro Vieira}
\affiliation{Max-Planck-Institut f\"ur Gravitationphysik, Albert-Einstein-Institut,  14476 Potsdam, Germany \& \\Centro de F\'\i sica do Porto,  Faculdade de Ci\^encias da Universidade do Porto, 4169-007 Porto, Portugal }

\begin{abstract}
We compute  the full dimension of Konishi operator in
planar  N=4 Super Yang-Mills (SYM) theory {\it for a wide range of couplings}, from weak to strong coupling regime,  and predict the subleading terms in its
strong coupling asymptotics. For this purpose we solve  numerically the integral
form of the AdS/CFT Y-system   equations for the exact energies of excited states
proposed by us and A.~Kozak.

\end{abstract}

\maketitle

\section{Introduction}

Four-dimensional Yang-Mills theories are at the heart of  modern high
energy physics, describing all fundamental interactions except gravity. Nevertheless,
in spite of considerable efforts during almost 40 years, we still don't have a satisfactory quantitative description of the most interesting YM theories, such as QCD, in the region of intermediate and strong couplings.
The low energy quantum
dynamics
of YM field is mostly known only from computer simulations of lattice YM theories.
A few important
exact results concerned the topological, BPS sectors of N=1,2 SYM were obtained.

Recently, when the hopes on  complete exact 4D solutions
in particular, for the quantities given by nontrivial 4D Feynman series
seemed to
start  waning, N=4 supersymmetric Yang-Mills theory  gave us  serious hopes for
a better understanding of the dynamics of strongly interacting 4D gauge theories.
 Due  to the  AdS/CFT correspondence \cite{Maldacena:1997re}, as well as to the quantum  integrability
discovered on both sides of this duality in the planar limit (when the  number of colors $N\to\infty$ with the 't~Hooft coupling $\lambda=g_{\rm YM}^2N$ fixed) \cite{Minahan:2002ve,Bena:2003wd,Beisert:2003tq,Kazakov:2004qf,Beisert:2005bm,Janik:2006dc,Staudacher:2004tk,Beisert:2005tm,Arutyunov:2004vx,Beisert:2006ib}, we acquire little by little  tools for the  study of
 the most important quantities in  N=4 SYM, such as the spectrum   of dimensions     \(\Delta(\lambda)\)  of local operators as functions of     \(\lambda\) -  the scale independent coupling constant in this superconformal  4D theory.   The weak coupling behaviour
 ($\lambda\to 0$) of         \(\Delta(\lambda)\)
is given by  Feynman perturbation theory. The dual string worldsheet \(\sigma\)-model
with the coupling $g=\sqrt\lambda/4\pi$
allows to find the strong coupling asymptotics of various dimensions.
Integrability allows us to connect the two regimes. In particular, the asymptotic Bethe ansatz (ABA) of \cite{Beisert:2006ez} gives us  the asymptotic spectrum of single trace operators containing an asymptotically large number of  elementary fields.

However, for short operators, such as Konishi operator \({\rm Tr} [D,Z]^2\)
\cite{Bianchi:2001cm}
\footnote{Here $Z$ is one of the complex scalars and $D$ is a covariant derivative in a light cone direction.}, the calculation of anomalous dimensions is still an interesting and important challenge.


Recently we proposed the Y-system  for the planar AdS/CFT \cite{Gromov:2009tv}, a set of functional equations   defining the
anomalous dimensions of {\it\ all} operators of planar N=4 SYM theory {\it at any coupling}. The  integral form of
the \(Y\)-system for
excited states in \(SL(2)\) sector, including the one corresponding to Konishi operator, was
presented in \cite{Gromov:2009bc}.
The integral equations for the BPS vacuum energy
were independently obtained in   \cite{Bombardelli:2009ns,Arutyunov:2009ur} by  the thermodynamical Bethe  ansatz (TBA) technique based on the dynamics of bound states   \cite{Zamolodchikov:1991et,Dorey:2006dq,Takahashi,KorepinBook,Bazhanov:1996aq,Dorey:1996re,Fioravanti:1996rz} (see also \cite{Beisert:2005tm,Arutyunov:2009zu}) of the mirror
theory \cite{Ambjorn:2005wa,Arutyunov:2007tc}.
The solutions of  the integral Y-system are also  solutions of the functional Y-system \cite{Bombardelli:2009ns,Gromov:2009bc,Arutyunov:2009ur,Frolov:2009in}.
\begin{figure}[t]
\includegraphics[width=80mm]{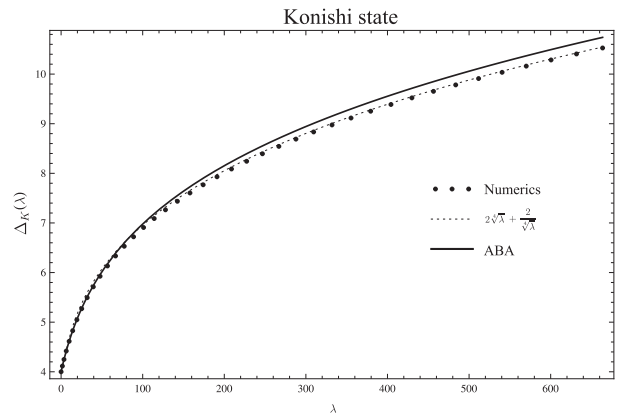}
\caption{
Numerical solution of exact finite size integral Y-system equations for the Konishi dimension \(\Delta_K(\lambda)\) in a wide range of 't~Hooft couplings
\(\lambda\),  compared to the asymptotic
Bethe ansatz curve and to the
predicted large \(\lambda\) asymptotics \(\Delta_K(\lambda)\simeq2\lambda^{1/4}+2/\lambda^{1/4}\)
obtained by fit.
} \la{Plots}
\end{figure}
The combination of   functional and  integral versions of the Y-system appears to be quite efficient to compute numerically the exact spectrum. In this work, we use the functional form of the Y-system to derive the large volume ($L$) \footnote{\(L\) is the number of \(Z\) fields in an operator in \(SL(2)\) sector.} asymptotic solution and then, departing from it, we solve the integral form of the Y-system iteratively.
As a demonstration of the power of our method, we calculate numerically  the dimension of Konishi operator  in a wide range  of the 't~Hooft coupling covering both the weak and strong coupling regimes.  The results appear to be quite satisfactory: we manage to compute  the dimension of Konishi operator  in the interval  of
couplings \(0\lesssim \lambda \lesssim700 \) and to confirm, within
the precision of our numerics,  all the existing data concerning this quantity: the perturbative results  \footnote{meaningful until the convergency radius \(g<g_c=1/4\); in this region our numerical
data are indistinguishable from the BAE for our accuracy. At  week coupling, an interesting numerical prediction could be the order  $g^{10}$. However our numerical error $\pm 0.001$ is larger than $(g_c)^{10}$ and therefore does not allow for such predictions.}  up to 4 loops (up to \(\lambda^{4}\) terms) \cite{Bajnok:2008bm,Bajnok:2008qj,Fiamberti:2008sh,Beisert:2007hz} and the large \(\lambda\) asymptotics \(2\lambda^{1/4}\) matches the prediction of  \cite{Gubser:1998bc}
for the lowest level of the string spectrum. Fitting our numerical data with $\lambda>60$ we find (with uncertainty in last digit)
\begin{equation}\Delta_K=2\lambda^{1/4}\left(1.0002+\frac{0.994}{\lambda^{1/2}}-
\frac{1.30}{\lambda}+\frac{3.1}{\lambda^{3/2}}+\dots
\right)\label{fit Konishi}\end{equation}
\begin{figure}[t]
\includegraphics[width=80mm]{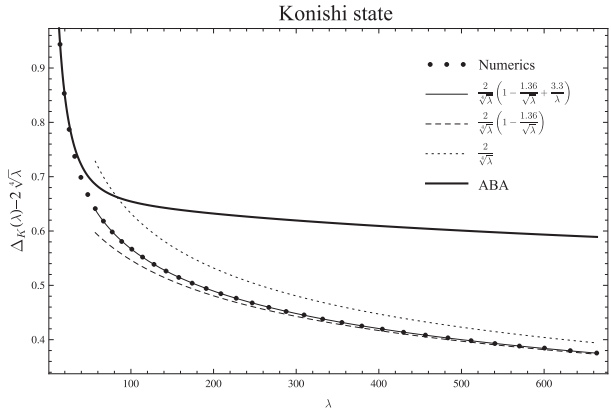}
\caption{ Plot of \(\Delta_K(\lambda)-2\lambda^{1/4}\) from the numerical data compared with the Bethe ansatz prediction and some fits. The fits in this plot are done assuming the asymptotics  \(\Delta_K(\lambda)=2\lambda^{1/4}+2/\lambda^{1/4}+\dots\).}\label{Plots2}
\end{figure}
The leading term reproduces indeed  the expected large $\lambda$ behavior
within our numerical precision.  It was also argued in
\cite{TseyRoi09} that the subleading  coefficient ought
to be integer (the next corrections could  be transcendental \cite{Tirziu:2008fk}). Indeed, our numerics seems to indicate that $\Delta_K=2\lambda^{1/4}+2/\lambda^{1/4}+\dots$
thus predicting the value of this integer!\footnote{Assuming the leading coefficient to be $2\lambda^{1/4}$ one gets
$1.001$ instead of $0.994$ for the subleading term, even closer to \(1\)
- the value predicted here.}
 We also obtained predictions for two further subleading corrections
(with a lower precision of course).


Our results are  represented in Figs.\ref{Plots} and \ref{Plots2}.

\section{Y-system functional and integral equations for AdS/CFT}
\begin{figure}[t]
\includegraphics[width=60mm]{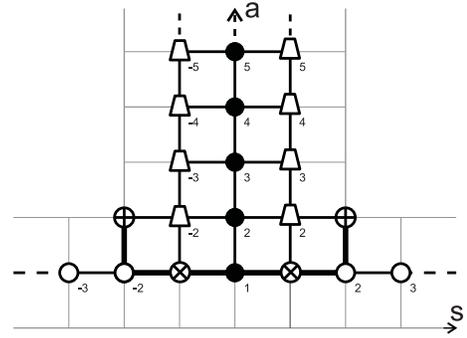}
\caption{\textbf{T}-shaped domain (\textbf{T}-hook) \cite{Kazakov:2007fy}. It defines the interactions between
\(Y\)'s in the AdS/CFT \(Y\)-system equations. }\label{FatHook}
\end{figure}

The  Y-system defining the spectrum of all local operators  in planar AdS/CFT correspondence reads \cite{Gromov:2009tv}
 \begin{equation}
\label{eq:Ysystem} \frac{Y_{a,s}^+ Y_{a,s}^-}{Y_{a+1,s}Y_{a-1,s}}
 =\frac{(1+Y_{a,s+1})(1+Y_{a,s-1})}{(1+Y_{a+1,s})(1+Y_{a-1,s})} \,.
\end{equation}
where the  functions \(Y_{a,s}(u)\) correspond  only to the nodes marked by \(\figb,\figf,\figF,\figp,\figm\) on Fig.\ref{FatHook} (we will use these
notations for \(Y\)-functions in what follows). The one particle energy at
infinite length  
$
\epsilon^{(a)}(u)= a+\frac{2ig}{x^{[-a]}}-\frac{2ig}{x^{[+a]}}
$
 is defined through the Zhukovski map
\(
x(u)+\frac{1}{x(u)}=\frac{u}{g} \, 
\)
and  \(f^{[\pm a]}\equiv f(u\pm ia/2),\quad f^\pm\equiv f(u\pm i/2)\) for any function \(f(u)\).
A solution
of \(Y\)-system with a given set of quantum numbers defines the energy of a state (or dimension of an operator in N=4 SYM) through the formula
\eqref{energy}
 where  the  Bethe roots $u_{j}$  are fixed by the exact Bethe ansatz equations $Y_{\figm_1}(u_j)=-1$.
In this paper we restrict ourselves to the integral form of the \(Y\)-system for the  \(SL(2) \)-excited states obtained in \cite{Gromov:2009bc}.
Furthermore we focus ourself for simplicity on the Konishi operator where we have
only  two   roots $u_{1}=-u_{2}$ which we can encode into the ``Baxter functions"  \(R^{(\pm)}(u)=(\mx(u)-x_1^\mp)(\mx(u)-x_2^\mp)\) and their complex
conjugates \( B(u)=\bar R(u)\) where $x_{1,2}^{\mp}=x(u_{1,2}\mp i/2)$ with the \textit{physical} choice of branches, such that $|x(u)|>1$, while for the free variable $\mx(u)$ we should use the \textit{mirror} kinematics which corresponds to the branches where $\im(x(u))>0$  \cite{Arutyunov:2007tc}. Unless it is explicitly said otherwise, we will be always using this latter choice in what  follows. With the mirror
choice,  $x(u)$ has a semi-infinite
cut for    $u\in (-\infty,-2g)\cup (2g,+\infty)$. The energy of the Konishi state is computed from
\beqa\label{energy}
\Delta_K=2+2\epsilon^{(1)}(u_{1})+\sum_{a=1}^\infty\int_{-\infty}^{\infty}\frac{du}{2\pi i}\partial_u \epsilon^{(a)} \log\left(1+Y_{\figm_a}\right)\,,
\eeqa
%
where the integral equations defining \(Y_{\figm_a}\) read \cite{Gromov:2009bc}
\beqa
\log Y_{\figF}\!\!&=&\!\!
K_{m-1}*\log(1+1/Y_{\figb_{m}})/(1+Y_{\figp_{m}})\nn\\
\!\!&+&\!\!{\cal R}^{(0m)}*\log(1+Y_{\figm_m})+\log\frac{-R^{(+)}}{R^{(-)}}\nn
\\
\log Y_{{\figp}_{n}}\!\!&=&\!\!{\cal M}_{nm}*\log(1+Y_{\figm_m})
-K_{n-1}\cut\log(1+Y_{\figF})
\label{PYR}\nn\\
\!\!&-&\!\!\nn
K_{n-1,m-1}*\log(1+Y_{{\figp}_{m}})+\log\frac{R_n^{(+)}{B_{n-2}^{(+)}}}{R_n^{(-)}B_{n-2}^{(-)}}
\\
\log Y_{{\figb}_{n}}\!\!&=&\!\!K_{n\!-\!1,m\!-\!1}\!\!*\!\log(1\!+\!1/Y_{{\figb}_{m}}\!)
\!+\!K_{n-1}\cut\log(1\!+\!Y_{\figF}\!)\nn \\
\log Y_{{\figm}_{n}}\!\!&=&\!\!
{\cal T}_{nm}*\log(1+Y_{\figm_m})+2{\cal R}^{(n0)}\cut\log(1+Y_{\figF})\nn
\\\!\!&+&\!\!
2{\cal N}_{nm}*\log(1+Y_{\figp_{ m}})+i\Phi_n\,\,.
\label{MIDN}\eeqa
We use here  the kernels and sources defined in \cite{Gromov:2009bc} and presented in the appendix for completeness. The  integrals in convolutions  $K*f=\int dv K(u,v)  f(v) $ go along the  real axis, but slightly  below
the poles in the terms involving \(Y_{{\figp}_{2}}\) (due to the last term
in the corresponding integral equation, see \cite{Gromov:2009bc}). The   convolutions $\cut$ should be understood in the sense of a B-cycle (see \cite{Gromov:2009bc}), e.g. ${\cal R}^{(n0)}\cut\log(1+Y_{\figF})$ stands for
\beq
\int_{-2g}^{2g} dv \left[ {\cal R}^{(n0)}\log(1+Y_{\figF})-{\cal B}^{(n0)}\log(1+1/Y_{\figf})
\right] \nn
\eeq
where \(\frac{1}{Y_{\figf}}\)
is the analytical continuation of \(Y_{\figF}\)   across the cut
   $u\in (-\infty,-2g)\cup(2g,+\infty)$. Summation over the repeated index $m$ is assumed  with $m\ge 2$ for $\figp_{\pm m}$ and $\figb_{\pm m}$, and $m\ge1$ for $\figm_m$.

A remarkable feature of all these equations, crucial for the success of our numerics and noticed in   \cite{Gromov:2009bc}, is the reality of all $Y$-functions in the integral equations.


\section{Exact Bethe equations}
The Y-system integral equations for the functions $Y_{a,s}$ need to be supplemented by the \textit{exact} Bethe equations
\( Y^{\ph}_{{\figm}_{1}}(u_{j})=-1\)  which reproduce the asymptotic Bethe equations of Beisert and Staudacher in the large $L$ limit \cite{Gromov:2009tv}.  To use this equation, we need  to analytically continue the last of eq.\eqref{MIDN} in the free variable $u$  from mirror to  physical plane and then  evaluate it at $u=u_1$. After some manipulations with contours
we find \beqa
&&\!\!\!\!\!\log Y^{\ph}_{{\figm}_{1}}(u_1)\!=\!
\tilde{\cal T}_{1m}\!*\!\log(1\!+\!Y_{\figm_m})\!+\!\log Z_{\figp_2}\!(u_1)
\label{BAE}\\
 &&\!\!\!\!\!+ i\Phi_{\ph}(u_1)
+2({\cal R}^{(10)}_{\ph,\mir}\cut K_{m-1}+K_{m-1}^-)*_{p.v.}\log(Z_{\figp_{ m}})\nn\\
&&\!\!\!\!\!+2{\cal R}^{(10)}_{\ph,\mir}\cut\log(1\!+\!Y_{\figF})\!-\!2\log \tfrac{u_1  -i/2 }{i}
\!-\! 2\sum_{j=1}^2\log\frac{\tfrac{1}{x_1^+}-x_j^+}{\tfrac{1}{x_1^-}-x_j^+}
\nn \eeqa
where $*_{p.v.}$. stands for principal value integration,    $\tilde{\cal T}_{1m}$
is the dressing phase in the physical kinematics while $i \Phi_\ph$ is the same as $i \Phi$ but evaluated in the physical region (see appendix).
We used  $1/Y_{\figf}(u_j\pm i/2)=0$ (following from (\ref{MIDN})) and denoted
$Z_{\figp_{m}}=(1+Y_{\figp_m})(1-1/m^2)(u^2-u_1^2)^{\delta_{m,2}}$
to isolate the poles in $Y_{\figp_2}$ at $u=u_j$ and to ensure decreasing asymptotics at large $u$ which is of course useful for the numerics. Finally, in contrast to $Y^{\ph}_{{\figm}_{1}}(u_1)$, the term $Z_{\figp_2}(u_1)$ is evaluated in mirror kinematics.


\section{ Numerics and its interpretation }
We solve the integral equations \eqref{MIDN} iteratively for the Konishi state. Namely, we find the $Y$-functions at a step $n$ by plugging the $Y$-functions of step $n-1$ into the r.h.s. of these equations. As the first step of the iterations we use the large \(L\), ABA solutions of the \(Y\)-system \cite{Gromov:2009tv}. At each step of  iterations we also  update the position of the Bethe roots by solving the exact Bethe equation of the previous section.
It is important to note that the r.h.s. of \eqref{BAE} remains, within our precision, purely imaginary in the process of iterations.

We should also truncate the infinite set of $Y$-functions to some finite
number. We explicitly iterate  the first $25$ $Y_{\figp_n}$'s and $25$ $Y_{\figb_n}$'s  at each  step. Then we also  extrapolate them to obtain  extra $40$ $Y_{\figp_n}$'s and $40$ $Y_{\figb_n}$'s   with higher numbers to replace the infinite sums in \eqref{MIDN} on the next step of iteration.  For $Y_{\figm_n}$ we can truncate the sums much earlier (typically $Y_{\figm_2}/Y_{\figm_1} \ll 0.1$): the first $5$ of them are largely enough for our  precision. Finally, the integrals along the real axis are  computed  along the stretch $(-X,X)$ with $X$ being a large cut-off. With these approximations, our absolute precision for the energy is around  $\pm 0.001$.

We solved the integral equations for a wide range of couplings    $0\lesssim\lambda \lesssim 700 \gg 1$
stretching  from the perturbative region up to this value, which is already a deep strong coupling regime \footnote{The expected
world-sheet
perturbation theory expansion at strong coupling is $\lambda^{1/4}(a_1+a_2/\sqrt\lambda+\dots)$. The first two coefficients
$a_i$ - are of order $1$ and we assume that all $a_i\sim 1$.
In our numerics $1/\sqrt{\lambda}\sim 0.04$.
The appropriate extrapolation procedure can easily increase the precision by one order
and we expect the absolute numerical error for our $a_2$ coefficient to be within $0.005$.}. We found no sign of any singularity and  the curve looks perfectly smooth. By this reason we believe that any new singularities, such as  those related to the L\"uscher \(\mu \)-terms, will not appear. This seems to be   the case perturbatively \cite{JanikPR} and our numerics seems to indicate that the integral form of Y-system we are solving describes exactly the full spectrum of planar N=4 SYM theory in $SL(2)$ sector. Although we cannot discard a possibility that some new singularities  could collide with the integration contours for even larger values of the 't~Hooft coupling (such extra singularities could be easily incorporated into our code), our numerical results suggest that this possibility is very unlikely.

With a precision of $0.001$ we can approximate the Konishi dimension in the range we considered  by
$ {\footnotesize \sqrt[4]{g^2+1}}\frac{ 252.93 h^4+384.74 h^3+674.13
  h^2+128.17 h+4}{35.67 h^4+51.43 h^3+99.71 h^2+29.29 h+1}
$ where $h=g^2/\sqrt{g^2+1}$ and $\lambda=16\pi^2 g^2$ (this function is just a shorthand for the data in Table I).

\section{Conclusions}

We presented here a numerical  method for solving  the Y-system for the Konishi dimension in planar N=4 SYM.
It opens the way to the systematic study of the spectrum
of many interesting states at any values
of the coupling. We also hope to simplify the Y-system in the future using the underlying Hirota integrable dynamics  and reducing it to a finite system of integral equations.

\section{Appendix}
We use: ${\cal P}^{(n)}(v)\equiv -\frac{1}{2\pi}\frac{d}{dv}\log \frac{\mx^{[+n]}_v}{\mx^{[-n]}_{v}} $, $K_{n}\equiv \frac{2n/\pi}{n^2+4u^2}$ and
\beqa
&&\!\!\!\! {\cal R}^{(nm)}(u,v)\equiv
\frac{\d_v}{2\pi i}\log \frac{\mx_u^{[+n]}-\mx_v^{[-m]}}{\mx_u^{[-n]}-\mx_v^{[+m]}}-\frac{1}{2i}{\cal
P}^{(m)}(v)\, ,\;\;\nn\\
&&\!\!\!\! {\cal B}^{(nm)}(u,v)\equiv
\frac{\d_v}{2\pi i}\log \frac{{1/\mx}_u^{[+n]}-\mx_v^{[-m]}}
{{1/\mx}_u^{[-n]}-\mx_v^{[+m]}}-
\frac{1}{2i}{\cal P}^{(m)}(v)\nn
\,,\\
&&\!\!\!\!{\cal M}_{nm}\equiv K_{n-1}\cut {\cal R}^{(0m)}+K^{\neq}_{n-1,m-1}\nn \,,\\
&&\!\!\!\!{\cal N}_{nm}\equiv{\cal R}^{(n0)}\cut K_{m-1}+K^{\neq}_{n-1,m-1} \nn \,, \\
&&\!\!\!\! K_{nm}\equiv
{\cal F}_n^u\!\circ\!{\cal F}_m^v\!\circ\! K_{2}(u-v),\,
K_{nm}^{\neq}\equiv{\cal F}_n^u\!\circ\!{\cal F}_m^v\!\circ\! K_{1}(u-v)\nn \,,
\end{eqnarray}
where the fusion operation \({\cal F}_n^u\circ A \equiv
\sum_{k=-\frac{n-1}{2}}^{\frac{n-1}{2}}A(u+ik)\). Finally, we also use a nice integral representation \cite{Gromov:2009bc} of the kernel \({\cal T}_{n,m}\) given by
\beqa
&&\!\!\!\! {\cal T}_{n,m}(u,v)
=
\la{reppp}
-{K}_{n,m}(u-v)-\frac{in}{2}\mathcal{P}^{(m)}(v) \,, \\
&&\!\!\!\! -2\sum_{a=1}^\infty\int
\!\!\left[{\cal B}^{(10)}_{n1}\left(u,w+i\tfrac{a}{2}\right){\cal B}_{1m}^{(01)}\left(w-i\tfrac{a}{2},v\right)
 + c.c.
 \right]dw\,\,, \nn
\eeqa
where  \({\cal B}^{(10)}_{nm}=
{\cal F}_n^u\circ{\cal F}_m^v\circ {\cal B}^{(10)}\). For the exact Bethe equations we should use this kernel in the mixed representation,
\beqa\nn
&&\!\!\!\!\tilde{\cal T}_{1m}=-\sum_{a=1}^{\infty}2{\cal B}_{\ph,\mir}^{(10)}(u,w+i\tfrac{a}{2})\!*\!{\cal B}_{\mir,\mir}^{(0m)}(w-i\tfrac{a}{2},v) \nn
\nn\\
&&\!\!\!\!-\sum_{a=1}^{\infty}2{\cal B}^{(10)}_{\ph,\mir}(u,w-i\tfrac{a}{2}-i0)\!*\!{\cal B}_{\mir,\mir}^{(0m)}(w+i\tfrac{a}{2},v)-{K}_{1m} \nn \,.
\eeqa
Finally the
source term \(\Phi_n(u)={\cal F}_n^u\circ\Phi(u)\), with
\beq
\Phi(u)=\frac{1}{i}\log\left[
\lb\frac{{\mx}^{-}}{{\mx}^{+}}\rb^{L+M}
S^2\frac{B^{(+)+}R^{(-)-}}{B^{(-)-}R^{(+)+}}\right]\;
\label{INT Y-sys}\eeq
where the BES \cite{Beisert:2006ez} dressing phase \(S(u)=\prod_{j=1}^2 \sigma\left(x^{[\pm]} ,x_j^{\pm} \right)\) should be taken in the mixed, mirror-physical
representation in the integral equations and in the physical-physical representation for $\Phi_{\rm ph}$ appearing in the exact Bethe equations.
We use the mirror-physical integral representation \cite{Gromov:2009bc}
\beqa
&&\log S=
\log\frac{B^{(-)+}}{B^{(+)+}} \\
&&\nn +\left({\cal B}^{(10)}(u,w+i0)*{\cal G}*\log\frac{R^{(+)}(u-i0)}{R^{(-)}(u-i0)}
+c.c.
 \right)
\eeqa
where \(
{\cal G}(u)\equiv
\frac{\d_{u}}{2\pi i}\log\frac{\Gamma(1-i u)}{\Gamma(1+i u)} 
\)
while for the physical-physical representation we can use the DHM integral representation \cite{Dorey:2007xn}.
Finally  \({\mathcal K}_{{\rm ph},{\rm mr}}(u,v)\) represents a kernel where we use the physical (mirror) branches for
 $u$ ($v$). For Konishi $L=2$ and it has
$M=2$ derivatives.\\ \\
{\bf Acknowledgments}

The work of NG was partly supported by the German Science Foundation (DFG) under
the Collaborative Research Center (SFB) 676 and RFFI project grant 06-02-16786.  The work  of VK was partly supported by  the ANR grants INT-AdS/CFT (ANR36ADSCSTZ)  and   GranMA (BLAN-08-1-313695) and the grant RFFI 08-02-00287.  We  thank G.~Arutyunov, S.~Frolov, R.~Hernandez, R.~Janik,  L.~Lipatov, T.~Lukowski, E.~Lopez,
 A.~Rej, R.~Roiban, R.~Suzuki,  M.~Staudacher, A.~Tseytlin and K.~Zarembo for interesting discussions.
NG and PV would like to thank CFP for hospitality and computational facilities.

%
%
%
\begin{table}[h!b!p!]
\caption{Numerical values of the Konishi anomalous dimension. The numerical error
should be within $\pm 0.001$}
\begin{tabular}{|c|r||c|r||c|r||c|r|}
  \hline
  $\frac{\sqrt{\lambda}}{4\pi}$ &$\Delta_K(\lambda)$ & $\frac{\sqrt{\lambda}}{4\pi}$ &$\Delta_K(\lambda)$&
  $\frac{\sqrt{\lambda}}{4\pi}$ &$\Delta_K(\lambda)$ & $\frac{\sqrt{\lambda}}{4\pi}$ &$\Delta_K(\lambda)$
  \\\hline
 0.00 & 4.0000 & 0.55 & 5.9251&  1.05 & 7.7689 &  1.60 & 9.3874\\
 0.05 & 4.0297 & 0.60 & 6.1330&  1.10 & 7.9301 &  1.65 & 9.5207\\
 0.10 & 4.1155 & 0.65 & 6.3342&  1.15 & 8.0876 &  1.70 & 9.6524 \\
 0.15 & 4.2488 & 0.70 & 6.5300&  1.20 & 8.2428 &  1.75 & 9.7823 \\
 0.20 & 4.4189 & 0.75 & 6.7207&  1.25 & 8.3943 &  1.80 & 9.9101\\
 0.25 & 4.6147 & 0.80 & 6.9079&  1.30 & 8.5431 &  1.85 & 10.0361\\
 0.30 & 4.8269 & 0.85 & 7.0885& 1.35 & 8.6895 & 1.90 & 10.1609 \\
 0.35 & 5.0476 & 0.90 & 7.2646 &   1.40 & 8.8343&    1.95 & 10.2847 \\
 0.40 & 5.2710 &  0.95 & 7.4366&  1.45 & 8.9752 &  2.00 & 10.4063 \\
 0.45 & 5.4934 &  1.00 & 7.6044&  1.50 & 9.1149 & 2.05 & 10.5265\\
 0.50 & 5.7120 &  & & 1.55 & 9.2519 & & \\  \hline
\end{tabular}
\end{table}


\end{document}